**Inorganic/inorganic composites through emulsion templating**

Tianhui Jiang[1], Shitong Zhou[1], Yinglun Hong[1], Erik Poloni[1], Eduardo Saiz[1], Florian Bouville[1]

[1]Centre for Advanced Structural Ceramics, Department of Materials, Imperial College London, London (UK)

**Abstract**

Inorganic/inorganic composites are found in multiple applications crucial for the energy transition, from nuclear reactor to energy storage devices. Their microstructures dictate a number of properties, such as mass transport or fracture resistance. There has been a multitude of process developed to control the microstructure of inorganic/inorganic composites, from powder mixing and the use of short or long fibre, to tape casting for laminates up to recently 3D printing. Here, we combined emulsions and slip casting into a simpler, broadly available, inexpensive processing platform that allow for *in situ* control of composite's microstructure that also enables complex 3D shaping. Emulsions are used to form droplets of controllable size of one inorganic phase into another, while slip casting enable 3D shaping of the final part. Our study shows that slip casting emulsions trigger a two-steps solvent removal that opens the possibility for conformal coating of porosity. By making magnetically responsive droplets, we form inorganic fibre inside an inorganic matrix *in situ* during slip casting, demonstrating a simple fabrication for long-fibre reinforced composites. We exemplify the potential of this processing platform by making strong and lightweight alumina scaffolds reinforced by a conformal zirconia coating and alumina with metallic iron fibres that displays work of fracture an order of magnitude higher than alumina.

**Introduction**

Inorganic/inorganic composites are at the heart of multiple technological materials for the energy transition, from ceramic/ceramic composites for aerospatial applications[1,2] or fusion nuclear reactors[3], to safer and more energy dense solid electrolyte for energy storage or fuel cells[4]. The microstructure of inorganic/inorganic composites varies in complexity and length scale[5,6].

Starting from the simplest structure, powder mixing can be used to form two phases' composites in which one is dispersed homogeneously within a continuous matrix. A variant of this idea is the formation of functionally graded microstructure by changing the spatial composition continuously in one direction[7]. The most successful application of this strategy is zirconia toughened alumina used in orthopaedic applications for their impressive mechanical properties, reaching higher than gigapascal strength and with a low sensitivity to slow crack growth[8]. These microstructures are now being used to develop the next generation of solid electrolyte and electrode for advanced energy storage[7]. The size of the microstructure in these composites is linked with the size of the particles and typically range from 100nm to a micron. Going one degree higher in terms of microstructural complexity, elongated 1D inorganic structure can be weaved into another inorganic matrix materials[1]. This concept is used in numerous applications, including the well-known ceramic matrix composites developed for safety critical uses, for instance in high temperature reactor, plane engines, or nuclear reactor[3]. Fully inorganic long- or short- fibre reinforced composites can also be produced with metallic fibre, for instance in concrete[9] or cermets[9]. The fibres range from 10 µm to millimetre size and tuning the properties of the interface between the fibre and matrix is crucial for the proper function of the composites[10]. Multiple processes can be used to fabricate long-fibre composites, from fibre textile infiltration, short fibre mixing, to more recently 3D printing to fabricate complex architecture *in situ*[11–13].

Finally, layered inorganic/inorganic composites have been fabricated using metal or ceramic alternating layers, with the goal of reaching high mechanical properties from developing crack arresting mechanisms or increasing the performance of capacitors[14–16]. The processes to obtain layered architectures is mostly based on tape casting, leading to layer from 20 μm to a millimetres, with some more recent development reaching 10 μm using templating with ice crystals[17–19]. All these composites are technically critical for numerous applications, and being able to add flexibility and capability in their production could facilitate or open new possibilities.

Emulsions are fundamental liquid mixture used in industrial applications to every kitchen[20]. They are formed when droplet of an immiscible liquid is stabilised by a surfactant into another. Because of this immiscibility, emulsions are thermodynamically unstable, and the presence of a surfactant allow the kinetic arrest of droplet coalescence[21,22]. The surfactant amount, stabilisation capacity, and mixing energy allow an extreme tunability of the droplet size from 300 nm up to a millimetre. The composition can be tuned to modulate the volume fraction of droplets from diluted emulsions up to value close or even slightly higher than the droplet packing limit around 60vol.% This high volume fraction of droplets can be achieve through the deformation of the droplets compared with hard sphere packing limit[23]. The applications of emulsions can be found in food product to incorporate fat into a water-based suspensions, to form porous structures for application in catalysis[24], filtration[25], heat exchange[26], biomedical scaffolds up to the fabrication of monodisperse droplets in microfluidic devices[27]. Up to now, this technique has only been used to fabricate porous materials as the droplet phase is left empty save the solvent used.

While the typical emulsion microstructure leads to homogeneously distributed droplets into another fluid matrix, advances in processing technology expand our spatial control of the droplets amount and configuration. The addition of magnetically responsive particles within the droplet phase opens the possibility of controlling the position and concentration of the droplets using spatially varying magnetic field intensity[28]. Besides, the droplets can form chains under a static magnetic field of sufficient intensity, which length and spatial configurations can be adjusted by the magnetic field intensity[26,29,30].

Compounding this spatial manipulation, emulsions templating with particles-laden droplets, and conventional ceramic processing would provide a new platform to fabricate complex inorganic/inorganic composites with controllable directional microstructure, complex 3D shaping, and composition tuning. These composites could present a variable anisotropy in their properties and could be used for increasing mass transport in fully solid electrode for energy storage device, produce catalytic support, all the way to enabling a simpler way to form long fibre inorganic composites *in situ* for high performance structural applications.

In this study, we use emulsions to fabricate complex microstructure in inorganic/inorganic composites. For the first time, we add inorganic particles inside both phases of the emulsion and disperse them through control of their surface chemistry. Because in these conditions the two phases have different density, we study and modify the rheology of the continuous phase to form a gel and avoid creaming or sedimentation issues. These emulsions are then shaped using slip casting, a simple, effective, and industrial scale process to make ceramic parts. This process leads to multi-stage solvent removal that have a strong impact on the final structure, which we study using microscopy and *in situ* process monitoring. Finally, using this newfound knowledge and process, we demonstrate how this technique can be used first to fabricate strong and lightweight alumina/zirconia composites. In the final example, we show how this new process can be used to form long metallic iron fibre inside an alumina ceramic matrix using magnetically assisted slip casting[31] which show improved fracture properties.

**Main text**

The microstructure of inorganic/inorganic composites can be controlled by using emulsion templating and combining this templating technique with magnetically responsive droplets open the fabrication to more complex architecture.

Our strategy to form inorganic/inorganic composites rely on a tweak on traditional emulsion-making formula: instead of using the immiscible liquid in the droplet to form porosity in the final microstructure upon drying, the droplet phase contains a stable dispersion of another type of inorganic particles (Figure 1a). The initial step is thus to formulate two stable and well-dispersed inorganic particles suspensions but with two immiscible solvents. A surfactant capable of stabilising the droplets is then added into the suspension that will be the continuous phase. The two suspensions are then mixed, with the mixing providing the necessary energy to break the suspension into droplets, that then coalesce up until there is enough surfactant to stabilise them.

We then use slip casting to concentrate and shape the emulsion into possible complex 3D shape in an inexpensive and fast way. Finally, we obtain a complex 3D shaped green body ready to be densified through pressing and sintering with a microstructure templated by the emulsion, i.e. with droplet-shaped assembly of material dispersed through another material (Figure 1b).

The addition of superparamagnetic iron oxide nano particles (SPIONs) into the droplet phase further expands the method as the emulsion droplet position and organisation can be controlled using low intensity magnetic field (Figure 1c). The presence of SPIONS in the droplet phase makes them susceptible to magnetic field, which can be used either to control their position throughout the sample, enabling the formation of graded microstructure for instance, but also to make the droplets self-assemble into chains. The formation of droplet chains leads to the formation of fibre-like structure in the sample that can be controlled by the direction of the magnetic field during casting, introducing anisotropic properties into the final sample.

While this strategy can be used for any composition of materials and couple of immiscible solvents, we choose two examples to demonstrate the potential of this technique: an alumina/zirconia composites for the isotropic emulsion templating and alumina/iron for the anisotropic emulsion templating.

The continuous phase is the same in both cases, with submicrometric alumina (Fig. S1) particles in water dispersed using poly-acrylic salt. Different amount of poly-vinyl alcohol (PVA) are added initial to dissolve the PVA before the dispersant and alumina are added and will serve as the emulsion surfactant. The droplet phases are based on organic solvents, decane for the zirconia-based suspension (Fig. S2) and light hydrocarbon oil for the iron oxide-based one. Both particle types are first coated with a layer of oleic acid to ensure a semi-steric dispersion in the non-polar solvents. The paraffin oil already contains around 10vol% of SPIONs also coated by oleic acid to ensure their dispersion.

The magnetic field during the slip casting is generated using two rare-Earth magnets with opposite polarity placed on each side of the slip casting mould, leading to the possibility of fabricating centimetre-size sample (Figure 1d). This strategy leads to the fabrication of microstructure-tailored inorganic/inorganic composites but the emulsions stability and successful casting is dictated by a delicate balance of forces controlled through the composition and characterised through rheology.

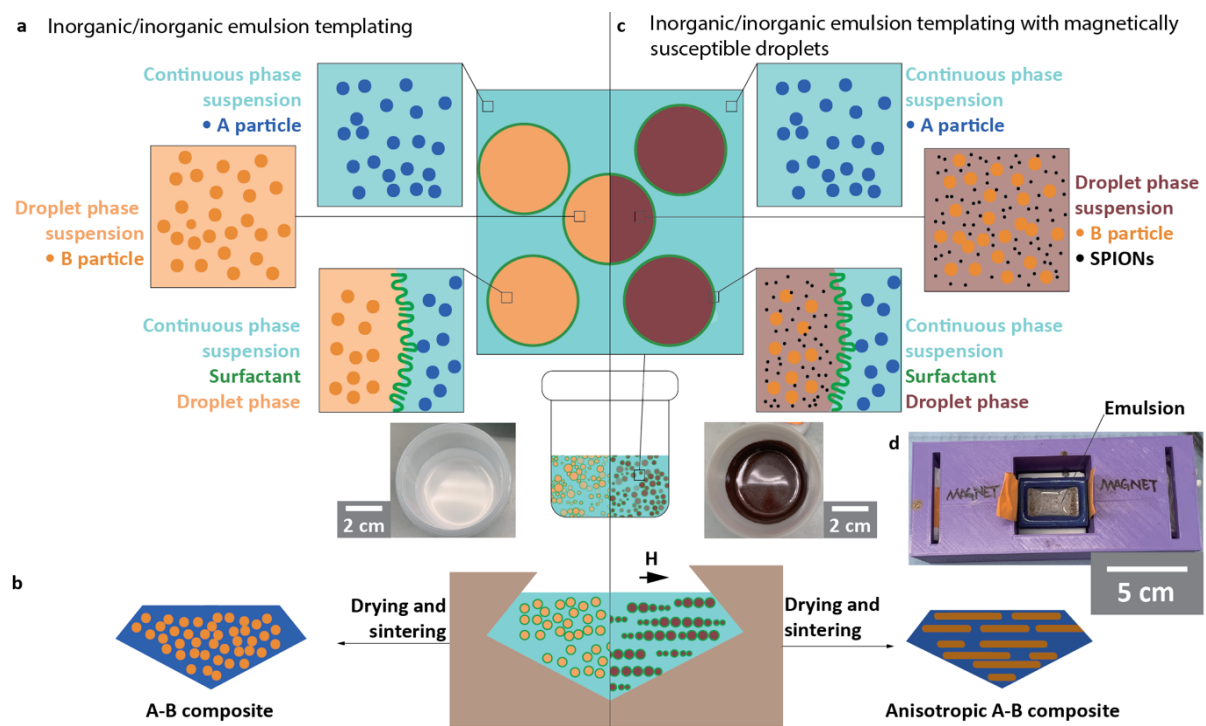

Figure 1: Fabrication of inorganic/inorganic composites through emulsion templating. a) Schematic of the emulsion composition for slip casting. Picture: resulting emulsion after mixing. b) Representation of the slip casting process with and without the magnetic field leading to different microstructure. c) Schematic of the emulsion composition for magnetically assisted slip casting. Picture: resulting emulsion after mixing. d) Picture of the slip casting setup using two rare-earth Neodymium magnets.

The stability and the structure of the emulsion can be controlled by both the surfactant and particles amount.

Once the separate suspensions based on immiscible solvents are ready, the mixing and the amount of the surfactant, in this case PVA, will dictate the stability of the emulsion during the casting. Our emulsions are even more susceptible to settling or creaming as conventional ones as the two liquid phases can present large differences in densities due to the presence of inorganic particles. We choose PVA as a surfactant first for simplicity and broad availability, but also for its capacity to increase the viscosity of the continuous phase. The initial idea was to use this increase in viscosity of the continuous phase to slow a possible settling/creaming of the droplets, however the role of PVA turned out to be even more important.

We used oscillatory rheology measurement to study the evolution of the viscosity of alumina water suspensions with 4wt.% and 7wt.% PVA with respect to the water (Figure 2a). Both suspensions present initially a liquid behaviour with the loss modulus G'' above the elastic modulus G', and the increase of the PVA from 4wt.% to 7wt.% increase the viscosity of the suspension by almost a factor of 5. However, the presence of PVA also triggers the gelation of the suspension over time, as proven by G' getting above G'' in both suspensions, attesting to a more solid-like behaviour of the suspensions after this point. The gelling time, taken as a first approximation as the time of the crossing between G' and G'', decreases with an increasing amount of PVA, going from more than 5 minutes to around a minute when the PVA amount increase from 4wt.% to 7wt.%.

During the next step, the two different water-based suspensions are emulsified with an oil-based suspension containing 20vol% of Zirconia. For clarity, we will use $\varphi$ as the volume fraction of solid particles in the oil phase and $\phi$ for the volume fraction of the oil phase, calculated as the volume of oil and volume of particles in it, with respect to the volume of the whole emulsion. The final emulsions are then cast over a gypsum mould that will remove the water from them. This drying method allows for a controlled solvent removal and consolidation of the powders in both phases, the kinetics of the solvent removal being dictated by the surface tension of the solvent and saturation of the gypsum[32,33]. The emulsions are then sintered to form alumina-zirconia composites with a microstructure that is directly templated by the emulsion. The two emulsions processes with the two different amounts of PVA led to different microstructures. A clear settling of the zirconia is visible with 4wt.% PVA, leading to a graded composition from the gypsum to the top of the sample (Figure 2b) while the emulsions made with 7wt.% of PVA presented a homogeneous microstructure throughout the sample thickness, with 10 microns diameters droplets of zirconia dispersed within an alumina continuous phase (Figure 2c). We suspect that the higher amount of PVA slows down the movements of the denser zirconia droplets within the alumina water-based continuous phase, while the gelling occurs to avoid further possibilities for the droplet to settle.

Based on this 7wt.% PVA alumina suspension, we can now study the effect of the varying amount of oil-based zirconia-suspension on the rheological behaviour of the emulsion. Because we want to form dense ceramic/ceramic composites, we study the behaviour of emulsions with the highest volume fraction of zirconia and alumina in both phases, corresponding to 20 vol% of zirconia in decane and 20 vol% of alumina in water. The viscosity as a function of the shear rate of the emulsion with an increasing addition of oil phase led to an increase in viscosity from the alumina suspension (Figure 2d). The behaviour of the emulsions follows first a similar trend as the alumina suspensions up to a volume fraction of oil phase $\phi$ of 0.50, with a viscosity decreasing more rapidly below 20s$^{-1}$ than above. However, this trend changes for a volume fraction of oil phase of $\phi = 0.60$, with the emulsion becoming more shear thinning. This behaviour is similar to the pure oil suspension, leading us to suspect a phase inversion at this point, with the oil suspension becoming the continuous phase over the water based one, thus dictating the rheological behaviour. This phase inversion was confirmed by SEM imaging of the resulting composites. The viscosity of the emulsions increases with addition of the more viscous oil-phase and an increase of the oil phase volume fraction above a threshold value led to the inversion continuous to the droplet phase, a behaviour akin to conventional emulsion[34].

We have established that the gelling of the emulsion is key to maintaining a homogeneous microstructure in the final composites, and the addition of the oil phase further shorten the gelation time, going from 72s in the suspensions down to 33s with the maximum volume fraction of oil phase. The addition of an increasing amount of oil-phase can be assimilated as an increase addition of particles in conventional ceramic suspension, with the interaction between the droplets dictating more and more of the rheological behaviour.

This increase in viscosity can be further quantified by plotting the viscosity of the suspension as function of the oil-phase volume fraction at two different shear rates, 20s$^{-1}$ and 100 s$^{-1}$. The addition of more oil-phase increases the viscosity of the emulsion following a power law, with the viscosity increasing faster the closer the volume fraction of oil-phase gets to a droplet packing limits. This behaviour, fitted by the Krieger-Dougherty model traditionally used with solid particles dispersed in a solvent, allow the extraction of this packing limit $\phi_{max}$.

$$\eta_r = \frac{\eta_{Emulsion}}{\eta_{Continuous\ Phase}} = \left(1 - \frac{\phi}{\phi_{max}}\right)^{-[\eta]\phi_{max}}$$

With $[\eta]$ being the intrinsic viscosity of the suspension by analogy with the formula developed by Einstein in dilute system[35]. With an oil-phase containing the maximum volume fraction of zirconia of $\varphi_Z = 0.20$, this packing limit is found to be $\phi_{max} = 0.60$ at a low shear rate. This packing limit is getting higher when the amount of zirconia in the droplet is decreased, with a packing limit of $\phi_{max} = 0.65$ reached for no zirconia. We hypothesised that the increasing amount of zirconia in the droplets changes the droplet rigidity and surface energy, leading to more agglomeration and lower packing. This trend is confirmed as the packing fraction extracted from the viscosity at higher shear rate, when the droplet interaction dictates less the viscosity, becomes similar for all volume fraction of zirconia in the oil. Finally, these plots allow us to establish the range within which the viscosity of the emulsion is low and thus castable, but also in which the droplets can be manipulated with a magnetic field later. For all emulsions, including the ones made with the iron oxide (Fig. S3) and containing SPIONs, the viscosity remains low below a volume fraction of $\phi = 0.4$ of oil-phase (Figure 2f,g).

The emulsions templated composites microstructure can now be characterised using SEM. Both the amount of zirconia and the amount oil-phase itself can be tuned and can lead to different droplets' sizes. First, we vary the amount of oil-phase at a constant $\varphi_Z = 0.20$ volume fraction of zirconia (Figure 2h) and measure how this influences the final size of the zirconia droplets within the alumina matrix. The zirconia droplets are hollow in all microstructures, which we assume is due to the presence of solvent in the droplets leading to the formation of porosity during drying, but this will be studied in more details later in the text. The droplets size distribution, measured using image analysis, is constant with the increase of oil-phase and spread from $D_{10}$~3µm to a $D_{90}$~15µm. This constant value reflects the stability obtained by the high PVA amount and gelling of the suspensions.

The microstructure of the emulsions obtained by varying the amount of zirconia at a constant volume fraction of oil-phase of $\phi = 0.50$ are represented in Figure 2i along with the droplets' size distributions. The addition of zirconia to the oil phase decreases the width of the droplet size distribution, with a $D_{90}$ decreasing from 25µm to 17µm when the zirconia content increases from 0 to 0.25, while the $D_{10}$ and $D_{50}$ stay relatively constant. This decrease in range of the droplet size points again toward a surface stabilising effect of the zirconia, even if more study would be necessary to study this quantitatively.

By tuning the amount of PVA, we obtained stable emulsions that lead to ceramic/ceramic composites directly templated by the emulsions. The droplets formed are all hollow, which can be directly linked with the behaviour of both solvents during the slip casting.

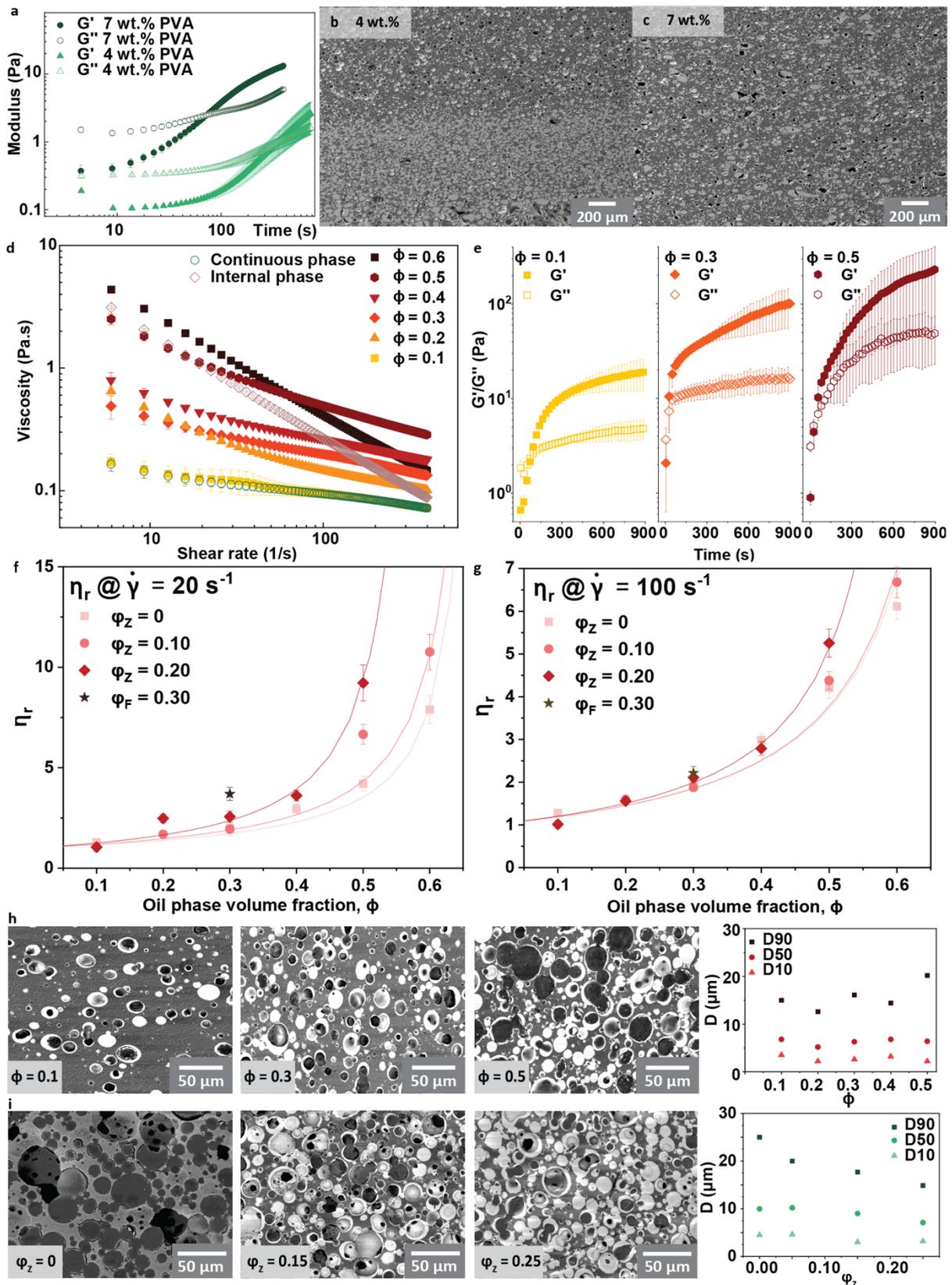

Figure 2. Control of the stability and viscosity of inorganic/inorganic emulsions through rheology modifier. a) Storage G' and loss G'' moduli as a function of time for 20vol%. alumina slurry in water with 4 and 7wt.% of PVA. SEM of the microstructure after drying and sintering of water-alumina/decane-zirconia emulsion with 4 b) and 7wt.% c) of PVA. d) Viscosity as a function of shear rate of water-alumina/decane-zirconia for different amount of decane-

zirconia phase $\phi$. e) Storage G' and loss G'' moduli as a function of time for different amount of decane-zirconia $\phi$. Relative viscosity of emulsions as a function of volume fraction of decane-zirconia $\phi$ with different fraction of zirconia $\varphi_Z$ or iron in decane $\varphi_F$ taken at a shear rate of f) 20s$^{-1}$ and g) 100s$^{-1}$. Microstructure of the water-alumina/decane-zirconia after slip casting and sintering for h) different amount of decane-zirconia $\phi$ at constant zirconia fraction in decane $\varphi_Z = 0.20$ and i) different amount of zirconia in decane $\varphi_Z$ at constant ratio of oil phase $\phi = 0.50$ and associate droplet size distribution after sintering.

The formation of hollow shell in the microstructure of emulsions templated composites can be explained by the formation of sequential two step slip casting.

The emulsion templated composites fabricated present an internal porosity regardless of the composition considered. While some porosity could be expected from the removal of the solvent from the droplets, the powder present in the droplet phase forms a thin shell surrounding the whole droplet. This observation as well as our curiosity as to how two immiscible solvents can be extracted from an assembly of powder led us to study the slip casting process further.

The addition of a water-soluble red dye, Rhodamine B, in the water-based alumina suspension and the use of black coloured SPIONs containing oil for the zirconia suspension helps visualise the extraction of the solvents by the gypsum mould over time. The side of the casting setup was covered with a glass slide to allow the direct visualisation of the emulsions and gypsum cross section using a portable microscope (Fig. S4). Images were recorded over time and analysed (Figure 3).

The extraction of solvent starts as soon as the emulsion is in contact with the gypsum and while the emulsion is coloured black due to the presence of SPIONs, the liquid extracted by the mould first is the pink-coloured water. The driving force for this solvent extraction is the capillary forces originating from the porous hydrophilic gypsum and lead to the concentration of the ceramic particles close to the mould. This increase in concentration forms a solid layer with solvent saturated jammed ceramic particles at the bottom of the emulsion that is growing along the height as the water keeps on getting extracted. This step has been studied in detail for ceramic suspension and the jammed layer thickness increase $t^{\frac{1}{2}}$ due to the increase jammed layer thickness and thus pressure drop that the solvent as to overcome[33]. By extension, it has been established that the depth of the solvent penetration in the gypsum also increases as $t^{\frac{1}{2}}$.

The images of taken during casting show that the water is extracted first, with only a few spots where the oil gets in the gypsum that do not spread further (Figure 3a). The water saturated gypsum is probably preventing the oil from being extracted at this stage. Plotting the profile of a single slice taken at the centre of the sample as a function of time allow to directly visualise the penetration depth of the water as function of time (Figure 3b). This confirms that the water depth $d_w(t)$ increases as $d_w(t) = k_w t^{\frac{1}{2}}$ in the emulsion, with $k_w = 0.0077 mm.s^{-\frac{1}{2}}$ a constant, but also shows that the oil starts to penetrate the mould after close to 3h of casting (10$^4$s, cf. Figure 3b). This sequential behaviour has also been observed in emulsion templated porous structure before[26].

The same recording is pursued at a slower image acquisition time of 200s to consider the slower oil penetration in the mould (Figure 3c). The images show a similar trend with the oil as the with water penetration in the mould, demonstrating that the mould is also able to use capillary forces to extract oil from a suspension. The depth of the oil penetration $d_o(t)$ follows a similar trend as well, with $d_o(t) = k_o\, t^{\frac{1}{2}} = 0.0029\, t^{\frac{1}{2}}$, however with rate estimated from $k_w/k_o$ 2.6-fold slower (Figure 3d). This decrease in penetration speed probably originates from the 6-times higher viscosity of the oil compared with water and the smaller difference in surface tension between oil and gypsum leading to lower capillary forces. These observations allow us to establish a clearer picture of the formation of the emulsion templated composite microstructure.

The water is first removed from the emulsion, leading to the jamming of the alumina particles saturated with water that will lock the oil in place in the droplet. Eventually, the water recess from the alumina body and so the oil can now be extracted (Figure 3e).

Because the oil must go through the alumina first and then the mould or evaporate at the surface, the porous alumina structure around the droplet is now the mould extracting the solvent in this slip casting. This leads to the formation of a thin jammed layer of zirconia forming on all the alumina surface, forming a conformal coating visible on all the microstructures (Figure 3e).

In summary, the formation of the emulsion templated composite via slip casting is possible and proceed by a two-step solvent removal, leading to the possibility of forming a continuous layer of materials inside the pores of a different materials. Now that the process is better understood, we can start using it to form both porous and dense composites with controlled microstructures.

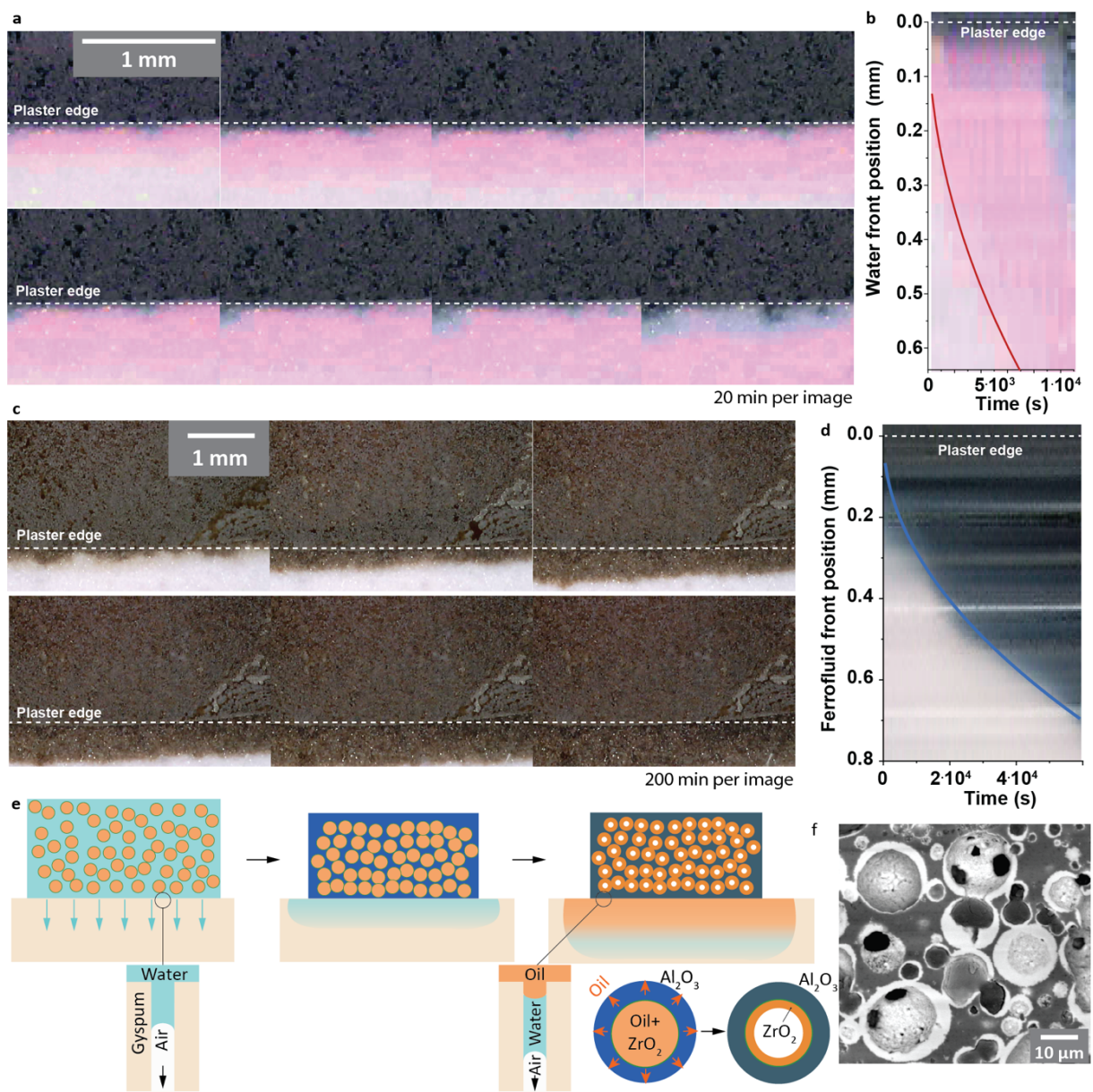

Figure 3 *In situ* optical characterisation of the solvent removal during slip casting of emulsion composites. Time series taken during the casting, a) first 160 minutes and c) from 160 minutes to 1360 minutes. Water contained rhodamine B as dye. Plot of the centre slice of the image as function of time superimposed with fit of the position of the solvent front in b) first 80 minutes with water and d) from 160 minutes to 1360 minutes with the oil based ferrofluid. e) Schematic representation of the two step slip casting mechanisms

We can use emulsion templating to control the microstructure of inorganic/inorganic composite either by using the two-step slip casting to form shells around pores in porous composites or by using a magnetically responsive oil droplet to form fibre reinforced composites *in situ*.

Starting from the initial emulsion composition developed with water-based alumina suspension as the continuous phase and zirconia containing oil as the droplet phase, we fabricated porous emulsion templated composites. The water-based suspension was kept constant as a matrix, with 20 vol.% alumina dispersed and 7 wt.% of PVA as a function of the water amount while the droplet phase was loaded with an increasing volume fraction of zirconia. The emulsion templated composites are slip cast and heat treated to burn the organic out and sinter the ceramic powder. The presence of a hollow core in the droplet due to the two-step slip casting leads to pores lined by a growing layer of dense zirconia as the volume fraction of zirconia is increased.

The porosity of the composites decreases almost linearly from 0.75 to 0.20 when the zirconia content in the oil phase increases from $\varphi_Z = 0$ to 0.20, while the oil phase volume fraction is kept constant throughout the whole series at $\phi = 0.20$ (Figure 4a). This linear decrease is expected as more and more of the droplet volume is replaced from oil to zirconia. However, the two-step slip casting allows for an additional control of the connectivity of the porosity compare with traditional emulsions. The amount of open porosity decreases when the amount of zirconia increases above 5 vol% in the oil phase, and almost no open porosity is kept at a volume fraction of zirconia of 25%. We explain this behaviour with the formation of the shell during the second slip casting step that once it reaches a certain thickness will start closing the connection between touching droplets. The sintered samples are tested in compression.

All compositions present a first failure followed by a more graceful damage increase; a behaviour typical of brittle foam (Figure 4b). The increase in zirconia content triggers an almost 9-fold increase in compressive strength, from 54 MPa to 480 MPa and decrease of the porosity from 73% down to 39% for a similar volume fraction of oil-phase in the initial suspension. While the presence of zirconia in the oil-phase decreases the porosity of the whole composite, it also prevents us the direct comparison of the addition of zirconia at equal porosity as the porosity reached with zirconia are too low to be reached without. We thus decided to compare the structural performance of the alumina zirconia porous composites with porous alumina fabricate by other methods but presenting a similar microstructure with spherical pores. The results are summarized in Figure 4c, and the alumina zirconia composites fabricated by emulsion templating present a compressive strength two times higher than alumina porous composites of similar porosity (37% and 39%). The presence of a zirconia shell thus seems to increase the mechanical resistance of the porous composites fabricated through emulsions templating.

Whereas this increase in structural properties could lead to further applications and studies, the possibility for microstructure control of this technique do not stop here. Using an oil phase containing both SPIONs and iron oxide particles, we show that we can form *in situ* long metallic fibre in a ceramic matrix using a magnetic field during slip casting. The presence of SPIONs and iron oxide makes the droplets susceptible to magnetic field. Multiple droplet arrangements can be observed in magnetically responsive droplet systems, but in most case an increase in magnetic field leads to the formation droplet chains as each droplet form a magnetic dipole under the effect of the macroscopic field. We confirm that this phenomenon occurs with our SPIONs/iron oxide oil phase droplet in a water suspension containing 7wt.% of PVA (Figure 4d). The presence of a magnetic field from a rare-Earth magnet put in proximity with the emulsion leads to the formation of droplet chains in the direction of the magnetic field.

Emulsion-templated composites containing alumina in the water phase, SPIONs and iron oxide in the oil phase are then slip cast under a static magnetic field (Figure 1d), then the iron oxide is reduced to metallic iron under reducing gas heat treatment at 600°C, before being compacted by CIP and finally sintered to almost full density in an inert atmosphere at 1450°C (see methods and Fig. S5).

The microstructure of the emulsion templated iron/alumina composites present the microstructure we were looking for, with long metallic fibre present within an alumina matrix (Figure 4e). The iron oxide containing droplet chains probably collapse into these continuous fibres during the cold pressing step.

While these microstructures could be interesting for more than structural properties enhancement depending on the compositions used, we illustrate their usefulness in increasing the fracture resistance of conventionally brittle alumina materials. The presence of metallic fibre in the alumina matrix changes completely the fracture behaviour of the composites, as confirmed during a fracture propagation test performed in three points bending setup (Figure 4f). The force-displacement shows an increase in the force after the first crack growth from the notch, with then a slow decrease of the force over a displacement of up to 0.3mm. The fracture toughness of the composites is $K_{IC} = 2.2 \pm 0.6\ MPa.\sqrt{m}$, a value that is lower than the value expected for pure alumina (around 3.5 $MPa.\sqrt{m}$ [36]). This decrease in initial toughness could be linked with the presence of residual stresses from the co-sintering of iron and alumina[11,12]. However, the work of fracture of the composites is $385 \pm 55\ J/m^2$, which is an order of magnitude higher than the one of typical alumina[37].

The reasons behind this increase in fracture resistance can be found by looking at the fracture propagation inside the composite during the test (Figure 4g). The crack can be seen stopping when it encounters a metallic fibre, then deflecting and growing around it, leading to crack bridging by the fibres. The bridging and pull-out of the fibre is even more apparent when looking at the composites cross-section post-fracture (Figure 4h), with evidence of pull-out and plastic deformation of the metallic fibre protruding from the alumina surface.

In summary, emulsions templating can be used to form complex microstructure in inorganic/inorganic composites that present improved structural properties.

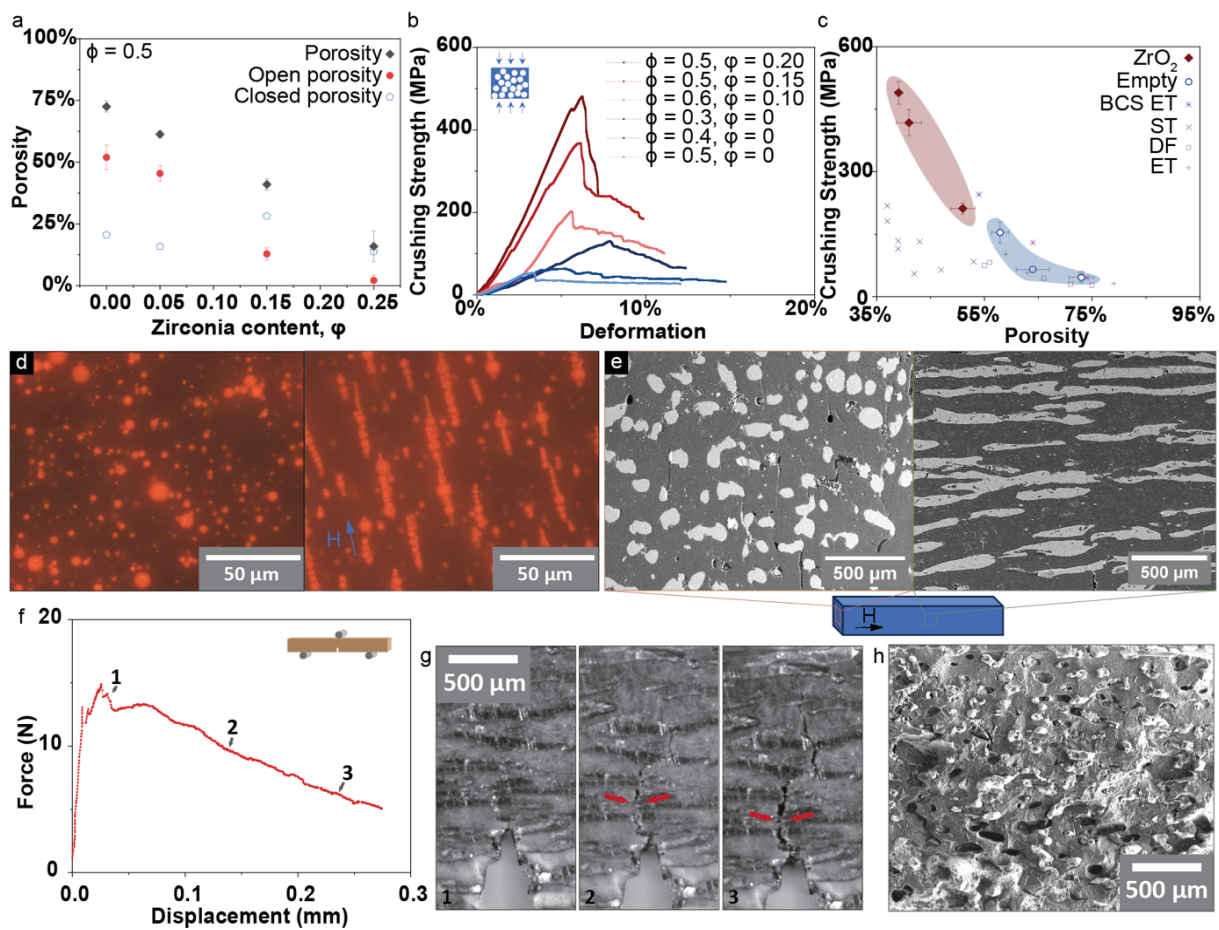

Figure 4. Porous and dense composites with controlled microstructure made by emulsion templating. a) Porosity as a function of the zirconia content at a constant initial volume fraction oil phase $\phi$=0.5. b) Stress-strain curves in compression of the porous alumina zirconia composite from emulsion templating. c) Crushing strength versus porosity for porous emulsion-based ceramic and our emulsion templated porous composites. Data obtained from references: BCS Emulsion Templating[22], Sacrificial templating[38], Direct foaming[39,40], Emulsion templating[41,42]. d) Optical microscopy images of iron oxides and SPIONS containing oil droplet in water 7wt.%. PVA solution with and without a static magnetic field applied. e) Microstructure of the alumina-iron composite made from magnetically templated emulsion after sintering. f) Typical force-displacement curves obtained for the alumina iron templated composites tested in Single Edge Notch Bending. g) Optical image taken during the fracture testing showing the crack propagation. h) SEM of the cross section of the composites after testing.

**Conclusions**

In conclusion, we demonstrated that by adding particles to both immiscible solvent in an emulsion we could produce inorganic/inorganic composites with complex microstructure. The control of the rheology of each phase and of the final emulsion through surface active additive is central to the success of the process. We found that PVA act both as a surfactant to stabilise the emulsion droplets and induce a temporal gelling that side-step any sedimentation/creaming issues. The median droplet size we obtain is on average lower than 10 μm but we envision that with different surfactant and higher energy mixing it should be possible to obtain smaller sizes as is the case with conventional emulsions. Using slip casting allows for a controlled solvent removal in the emulsion and using *in situ* observation we unveiled a two-step solvent removal that leads to conformal coating of the inner surface of the droplets by the particles present in the oil phase. The final composites display improved properties directly linked with the microstructural control enable by the emulsion templating process. Porous alumina with zirconia coated pores present higher crushing strength than porous alumina. Using magnetic field to form *in situ* iron fibre within an alumina matrix leads to an increase fracture resistance due to the crack bridging and fibre strain hardening. We envision that inorganic/inorganic composites fabricated by emulsion templating can bring further breakthrough in the fabrication of energy storage device, catalytic support, or long fibre inorganic composites.

**Materials and methods**

**1.1 Materials**

$Al_2O_3$, alumina powder (SMA6, Baikowski, $d_{50}$=0.2μm), ZrO2, zirconia powder (TZ-3Y-E, Tosoh), $Fe_2O_3$, iron (III) oxide powder (<5 μm, ≥99%, Sigma-Aldrich), polyvinyl alcohol (PVA, 98-99% hydrolysed, average M.W. 11,000-31,000, Alfa Aesar), Dolapix CA (Zschimmer & Schwarz), oleic acid (Fluorochem Ltd.), decane (≥99%, Sigma-Aldrich) or oil-based ferrofluid (EFH-1, Ferrotec Corp.).

**1.2 Water-based Slurry Preparation**

PVA was dissolved in deionized water at 80°C to prepare aqueous solutions at 4 and 7 wt.%. $Al_2O_3$ powder was sieved with a 100 μm mesh break the large agglomerates present in the received powder. Then, the sieved powder was mixed with the PVA solution in a 150 ml HDPE container and 0.5 wt.% Dolapix CA (with respect to weight of powder) was used as dispersant. A planetary centrifugal mixer (THINKY ARE-250, Thinky Corp.) was used for mixing at 2000 rpm for 2 min, repeated for 3 times, followed by a defoaming step (2200 rpm for 10 min) to get a smooth and homogeneous slurry.

**1.3 Oil-based Slurry Preparation**

$ZrO_2$ and $Fe_2O_3$ powder was de-agglomerated in an ethanol suspension by 6-hour ball milling using a shaker mixer (TURBULA Type T2C, WAB). The ethanol suspension contains 1.9 wt.% or 2.0 wt.% to weight of $ZrO_2$ and $Fe_2O_3$ powder respectively of oleic acid, 10 vol.% powder and zirconia milling media (ball/powder weight ratio of 3.5/1). By removing the ethanol using a rotary evaporator (Rotavapor R-300, Buchi Ltd.), oleic acid-coated powder can be obtained after drying in convection oven at 70°C overnight. $ZrO_2$ or $Fe_2O_3$ powder with precoated oleic acid was weighed and dispersed in decane or oil-based ferrofluid using Thinky mixer (2000 rpm, 2 min, 3 times).

**1.4 Emulsion Preparation and Slip Casting**

Next, the emulsification process was carried out using Thinky mixer (2000 rpm, 2 min, 3 times). The emulsion was slip cast in a cylindrical mould put on a plaster. To prevent the emulsion from sticking to the mould, silicon oil was applied on the walls. It could also help a more uniform shrinkage of the emulsion and reduce cracks formation at the surface or inside the green body. To aid easy release of the green body from the plaster, a piece of filter paper was placed between the emulsion and the plaster, preventing the green body from sticking to the plaster plate and avoiding fracture during removal.

## 1.5 Magnetically Assisted Slip Casting

To facilitate magnetic templating for emulsions containing a magnetically responsive droplet phase, a static magnetic field was employed during the casting process. The magnetic field was generated by two block neodymium magnets (H x L x T = 50.8 x 50.8 x 25.4 mm; $(BH)_{max}$ = 306 kJ m$^{-3}$; magnetic flux density, $B_m$ = 450 mT, Magnet Sales, UK). A 3D-printed setup was designed to keep the magnets apart, leaving space for a plaster plate for slip casting.

A silicone casting mould was used. It has internal dimensions of 14 x 27 x 20 mm, and wall thickness of 4mm. The mould corners were curved, and its walls were coated with silicon oil to avoid green body breakage due to uneven shrinkage and adhesion with the mould. The emulsion was cast at the centre of the plaster for a strong and uniform magnetic field, and the plaster plates were cast to be 45 x 45 x 20 mm, allowing the emulsion to be cast in the centre of the magnetic field.

## 1.6 Drying, Reducing and Sintering

The green bodies were left to air-dry at room temperature for at least 96 hours. For $Al_2O_3$ - $ZrO_2$ composites, a lift furnace was used for sintering in air. Pre-sintering at 500 °C for 2 hours was done for debinding, followed by a 1-hour sintering at 1550 °C, with ramp rates of 2 °C/min and 5 °C/min respectively.

For magnetically assisted slip cast samples with $Fe_2O_3$, the green bodies were first reduced at 600 °C in a 10% H2 / Ar gas for 36 hours. Then, cold isostatic pressing (CIP, TCH Instrument Co.) was applied to green bodies at 350 MPa for 5 minutes to close pores and improve densification. Sintering was carried out in a high-temperature vacuum furnace with a tungsten heating element and 10% $H_2$ / Ar gas at 1450 °C for 1.5 hours, with a ramp rate of 5 °C/min.

## 2. Characterisation

Images taken using scanning electron microscope (Auriga, Zeiss, Germany and JSM 6010LA, JEOL, Japan) were used for measuring droplet phase size distribution in image analysis (ImageJ, NIH, USA).

Densities were measured and calculated using Archimedes' method by employing the theoretical density calculated based on emulsion compositions.

## 2.1 Rheological characterisation

The Discovery Hybrid Rheometer (DHR-1, TA Instrument, UK) and a 40-mm steel parallel plate geometry was used for measuring the rheological behaviours of slurries and emulsions. The testing gap was 500 μm and a solvent trap was used to reduce the effect of evaporation.

Flow ramp tests were carried out with shear rate range from 0.5 to 400 s$^{-1}$ at 20 °C. Oscillation time sweep tests were carried out with fixed oscillation at 1.0 Pa stress, 0.5 Hz frequency for 900 s at 20 °C.

## 2.2 Time-lapse filming

A digital microscope (Dino-Lite Premier AM7013MZT) with DinoCapture 2.0 software were used to record the process in time lapse mode once a layer of emulsion was slip casted on plaster. To better distinguish water based continuous phase and oil-based droplet phase, water soluble Rhodamine B (pink colour) was added during preparation of water phase suspension and ferrofluid with brown colour was applied as medium of droplet phase. The image sequence recorded by time-lapse filming was cropped and rearranged as a function of time using Reslice plugin in ImageJ software[43].

## 2.3 Mechanical Tests

The compressive strength, $\sigma_c$ of porous samples was measured by performing compression tests using the Zwick universal test machine (Z010, ZwickRoell) on cubic samples with dimensions of 4.5 × 4.5 × 4.5 mm, with a loading speed of 1 mm·min$^{-1}$. The compressive strength, $\sigma_c$ and strain, $\varepsilon$ can be calculated using following equations:

$$\sigma_c = \frac{F}{A} \qquad \text{Equation 2.1}$$

$$\varepsilon = \frac{D}{d} \qquad \text{Equation 2.2}$$

where D is displacement, d is sample thickness, A is area of the face under loading and F is applied force.

For three-point bending test operated on SENB samples, the supporting span of the fixture is 20 mm, while the loading speed is 0.06 mm·min$^{-1}$. The SENB samples were prepared with length of 25 mm, thickness, 2.5 < $d$ < 3.0 mm, and width $\frac{d}{2} < b < d$. The notch was firstly cut using a diamond wafering blade (thickness of 0.25 mm), followed by extending the notch using a razor blade and oil-based diamond suspension (1 μm) for 1 hour. The full length of the notched tip, $a_n$, is prepared to be: $\frac{d}{3} < a_n < \frac{d}{2}$.

A digital camera (Oryx 10GigE, Teledyne FLIR) combined with a telecentric lens (VS-LTC3.3–45/FS, VS Technology) was applied to record the process at a frame rate of 2 Hz for correction of real displacement during test with a resolution of 1 μm/pixel and imaging field size of 6.46 × 4.85 mm$^2$. With image processing using 3D drift correction plugin in ImageJ, image series can be converted into real displacement during SENB test, and contrast adjustment can be applied to track the propagation of crack extension.

The fracture toughness, $K_{IC}$, during three-point SENB test can be calculated using (ASTM-1820-11):

$$K_{IC} = \frac{3FL_s}{2bd^{\frac{3}{2}}} \cdot \left(\frac{a_n}{d}\right)^{\frac{1}{2}} \times \frac{1.99 - \frac{a_n}{d} \cdot \left(1 - \frac{a_n}{d}\right) \cdot \left[2.15 - 3.93\frac{a_n}{d} + 2.7\left(\frac{a_n}{d}\right)^2\right]}{\left(1 + \frac{2a_n}{d}\right) \cdot \left(1 - \frac{a_n}{d}\right)^{\frac{3}{2}}} \qquad \text{Equation 2.3}$$

F is the applied force at which F – d curve diverges from a linear relation.

The total energy of fracture process $W_f$ can be calculated from the force-displacement curve, hence work of fracture, $\gamma_{WOF}$, can be calculated from **Error! Reference source not found.**:

$$W_f = \int F d\delta \qquad \text{Equation 2.4}$$

$$\gamma_{WOF} = \frac{W_f}{2A_f} \qquad \text{Equation 2.5}$$

where $d\delta$ is change of displacement and $A_f = b(d - a_n)$ is area of fractured surface.

## Acknowledgements

This work was supported by the EPSRC Program Manufacture Using Advanced Powder Processes (MAPP)EP/P006566. F.B. acknowledges support from the European Research Council Starting Grant [H2020-ERC-STG grant agreement n°948336] SSTEEL. S.Z. acknowledges financial assistance from the China Scholarship Council (No. 201806120001)

## References

1. Bansal, N. & Lamon, J. L. *Ceramic Matrix Composites*. (Wiley-VCH, 2014).
2. Padture, N. P. Advanced structural ceramics in aerospace propulsion. *Nat Mater* **15**, 804–809 (2016).
3. Eswarappa Prameela, S. *et al.* Materials for extreme environments. *Nat Rev Mater* **8**, 81–88 (2023).
4. Kalnaus, S., Dudney, N. J., Westover, A. S., Herbert, E. & Hackney, S. Solid-state batteries: The critical role of mechanics. *Science (1979)* **381**, (2023).
5. Newnham, R. E., Skinner, D. P. & Cross, L. E. Connectivity and piezoelectric-pyroelectric composites. *Mater Res Bull* **13**, 525–536 (1978).
6. Kokkinis, D., Bouville, F. & Studart, André. R. 3D Printing of Materials with Tunable Failure via Bioinspired Mechanical Gradients. *Adv Mater* **1705808**, 1705808 (2018).
7. Naebe, M. & Shirvanimoghaddam, K. Functionally graded materials: A review of fabrication and properties. *Appl Mater Today* **5**, 223–245 (2016).
8. Chevalier, J. *et al.* Forty years after the promise of «ceramic steel?»: Zirconia-based composites with a metal-like mechanical behavior. *Journal of the American Ceramic Society* **103**, 1482–1513 (2020).
9. Barros, J. A. O., Cunha, V. M. C. F., Ribeiro, A. F. & Antunes, J. A. B. Post-cracking behaviour of steel fibre reinforced concrete. *Materials and Structures/Materiaux et Constructions* **38**, 47–56 (2005).
10. Rebillat, F., Lamon, J. & Guette, A. Concept of a strong interface applied to SiC/SiC composites with a BN interphase. *Acta Mater* **48**, 4609–4618 (2000).
11. Cai, Q. *et al.* 3D-printing of ceramic filaments with ductile metallic cores. *Mater Des* **225**, 111463 (2023).
12. Zhou, S., Iuliia, S., Zhang, X., Withers, J. & Bouville, F. Article Embedded 3D printing of microstructured multi-material composites Embedded 3D printing of microstructured multi-material composites. *Matter* 1–17 (2024) doi:10.1016/j.matt.2023.10.031.
13. Wilkerson, R. P. *et al.* A study of size effects in bioinspired, "nacre-like", metal-compliant-phase (nickel-alumina) coextruded ceramics. *Acta Mater* **148**, 147–155 (2018).
14. Bloyer, D., Ritchie, R. O. & Rao, K. Fracture toughness and R-curve behavior of laminated brittle-matrix composites. *Metallurgical and Materials Transactions …* **29**, (1998).
15. Okuma, G. *et al.* Microstructural evolution of electrodes in sintering of multi-layer ceramic capacitors (MLCC) observed by synchrotron X-ray nano-CT. *Acta Mater* **206**, 116605 (2021).
16. Bermejo, R. "Toward seashells under stress": Bioinspired concepts to design tough layered ceramic composites. *J Eur Ceram Soc* (2017) doi:10.1016/j.jeurceramsoc.2017.04.041.
17. Noirjean, C., Marcellini, M., Deville, S., Kodger, T. E. & Monteux, C. Dynamics and ordering of weakly Brownian particles in directional drying. *Phys Rev Mater* **1**, 065601 (2017).
18. Poloni, E. *et al.* Tough metal-ceramic composites with multifunctional nacre-like architecture. *Sci Rep* **11**, 1–12 (2021).
19. Wat, A. *et al.* Bioinspired Nacre-Like Alumina with a Metallic Nickel Compliant Phase Fabricated by Spark-Plasma Sintering. *Small* **1900573**, 1–7 (2019).
20. McClements, D. J. Critical review of techniques and methodologies for characterization of emulsion stability. *Crit Rev Food Sci Nutr* **47**, 611–649 (2007).

21. Studart, André. R., Gonzenbach, U. T., Tervoort, E. & Gauckler, L. J. Processing Routes to Macroporous Ceramics: A Review. *Journal of the American Ceramic Society* **89**, 1771–1789 (2006).
22. Garcia-Tunon, E. *et al.* Designing smart particles for the assembly of complex macroscopic structures. *Angew Chem Int Ed Engl* **52**, 7805–8 (2013).
23. Villone, M. M. & Maffettone, P. L. Dynamics, rheology, and applications of elastic deformable particle suspensions: a review. *Rheol Acta* **58**, 109–130 (2019).
24. Stein, A., Melde, B. J. & Schroden, R. C. Hybrid inorganic-organic mesoporous silicates-nanoscopic reactors coming of age. *Advanced Materials* **12**, 1403–1419 (2000).
25. Yang, X. Y. *et al.* Hierarchically porous materials: Synthesis strategies and structure design. *Chem Soc Rev* **46**, 481–558 (2017).
26. Ammann, J., Ruch, P., Michel, B. & Studart, André. R. High-Power Adsorption Heat Pumps Using Magnetically Aligned Zeolite Structures. *ACS Appl Mater Interfaces* **11**, 24037–24046 (2019).
27. Shah, R. K. *et al.* Designer emulsions using microfluidics. *Materials Today* **11**, 18–27 (2008).
28. Demirörs, A. F., Pillai, P. P., Kowalczyk, B. & Grzybowski, B. a. Colloidal assembly directed by virtual magnetic moulds. *Nature* **2**, 3–7 (2013).
29. Ivey, M., Liu, J., Zhu, Y. & Cutillas, S. Magnetic-field-induced structural transitions in a ferrofluid emulsion. *Phys Rev E Stat Nonlin Soft Matter Phys* **63**, 1–11 (2001).
30. Sander, J. S., Erb, R. M., Li, L., Gurijala, A. & Chiang, Y.-M. High-performance battery electrodes via magnetic templating. *Nat Energy* **1**, 16099 (2016).
31. Le Ferrand, H., Bouville, F., Niebel, T. P. & Studart, André. R. Magnetically assisted slip casting of bioinspired heterogeneous composites. *Nat Mater* **14**, 1172–1179 (2015).
32. ADCOCK, D. S. & McDOWALL, I. C. The Mechanism of Filter Pressing and Slip Casting. *Journal of the American Ceramic Society* **40**, 355–360 (1957).
33. TILLER, F. M. & TSAI, C.-D. Theory of Filtration of Ceramics: I, Slip Casting. *Journal of the American Ceramic Society* **69**, 882–887 (1986).
34. Maffi, J. M., Meira, G. R. & Estenoz, D. A. Mechanisms and conditions that affect phase inversion processes: A review. *Canadian Journal of Chemical Engineering* **99**, 178–208 (2021).
35. Mueller, S. *et al.* The rheology of suspensions of solid particles. *Proceedings of the Royal Society A: Mathematical, Physical and Engineering Sciences* **466**, 1201–1228 (2010).
36. Evans, A. G. Perspective on the Development of High-Toughness Ceramics. *Journal of the American Ceramic Society* **73**, 187–206 (1990).
37. RICE, R. W., FREIMAN, S. W. & BECHER, P. F. Grain-Size Dependence of Fracture Energy in Ceramics: I, Experiment. *Journal of the American Ceramic Society* **64**, 345–350 (1981).
38. RYSHKEWITCH, E. Compression Strength of Porous Sintered Alumina and Zirconia. *Journal of the American Ceramic Society* **36**, 65–68 (1953).
39. Dhara, S. & Bhargava, P. Influence of slurry characteristics on porosity and mechanical properties of alumina foams. *Int J Appl Ceram Technol* **3**, 382–392 (2006).
40. Hüppmeier, J. *et al.* Oxygen feed membranes in autothermal steam-reformers - A robust temperature control. *Fuel* **89**, 1257–1264 (2010).

41. Barg, S., Binks, B. P., Wang, H., Koch, D. & Grathwohl, G. Cellular ceramics from emulsified suspensions of mixed particles. *Journal of Porous Materials* **19**, 859–867 (2012).
42. Cesconeto, F. R. & Frade, J. R. Cellular ceramics by slip casting of emulsified suspensions. *J Eur Ceram Soc* **40**, 4949–4954 (2020).
43. Schindelin, J. *et al.* Fiji: an open source platform for biological image analysis. *Nat Methods* **9**, 676–682 (2012).

## Supplementary information

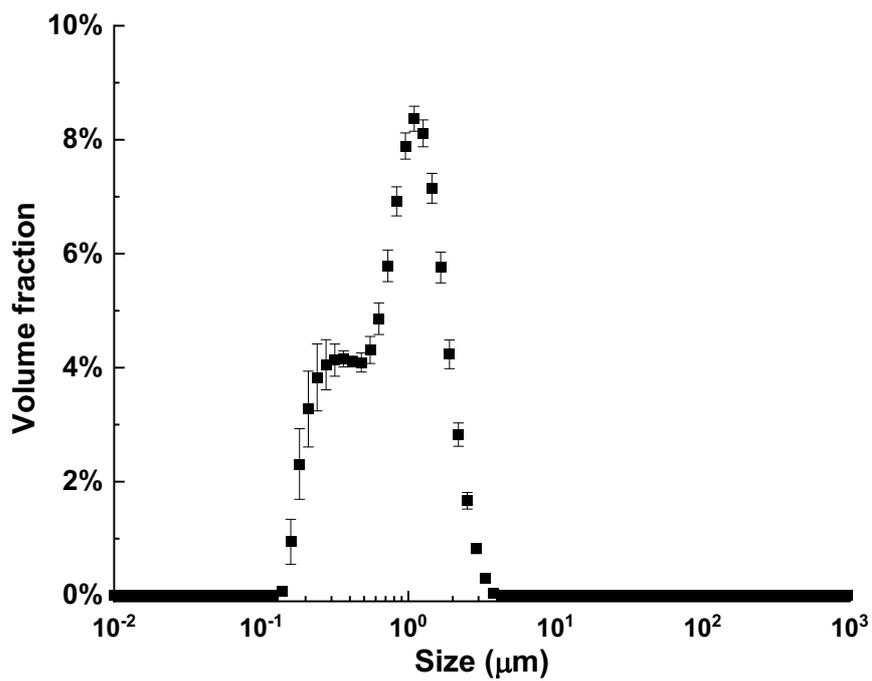

*Figure S1. Size distribution of sieved alumina particles dispersed in 7 wt.% PVA water solution with 0.5 wt.% Dolapix CA as dispersant, assessed by laser diffraction*

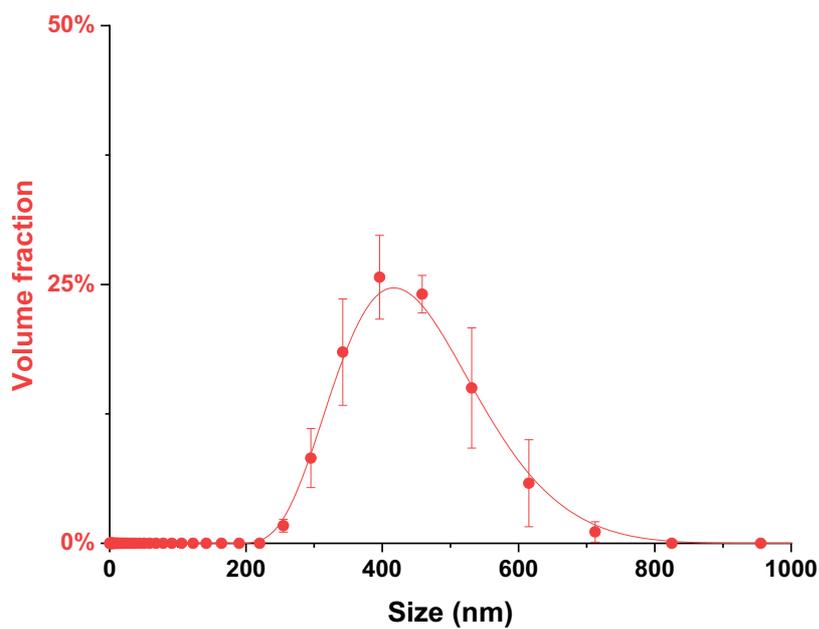

*Figure S2. Size distribution of zirconia particles in decane (precoated with 1.9 wt.% oleic acid), measured by dynamic light scattering.*

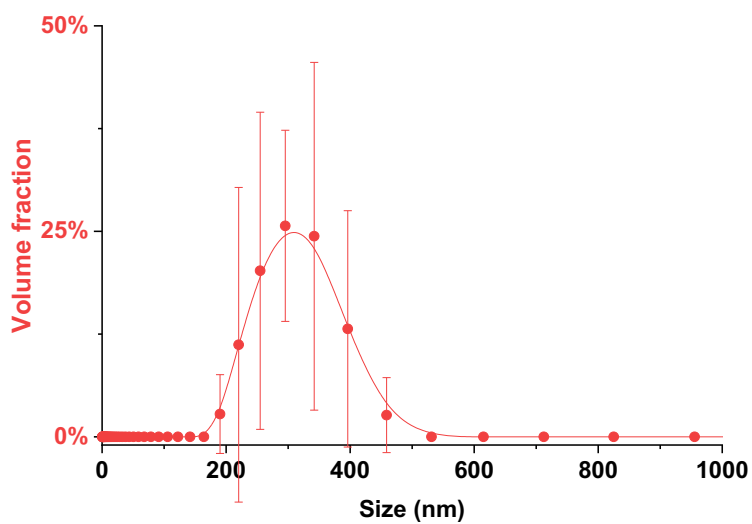

*Figure S3. Size distribution (by volume) of iron (III) oxide particles dispersed in decane (precoated with 2.0 wt.% oleic acid) ), measured by dynamic light scattering.*

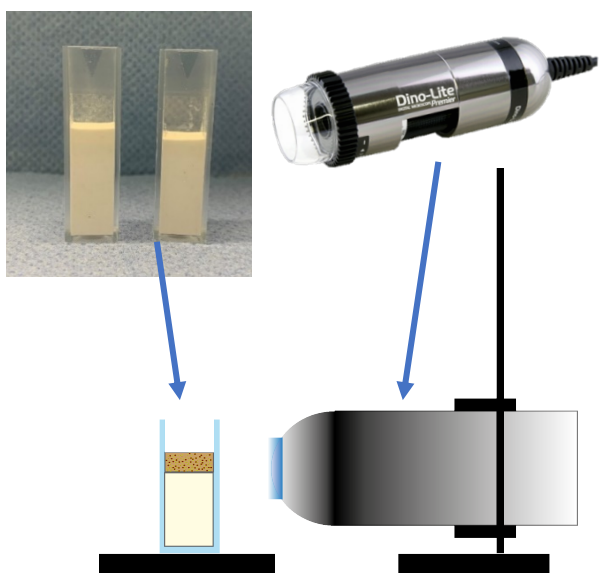

*Figure S4. graphic illustration of setup used for time-lapse filming to record emulsion solidified on plaster.*

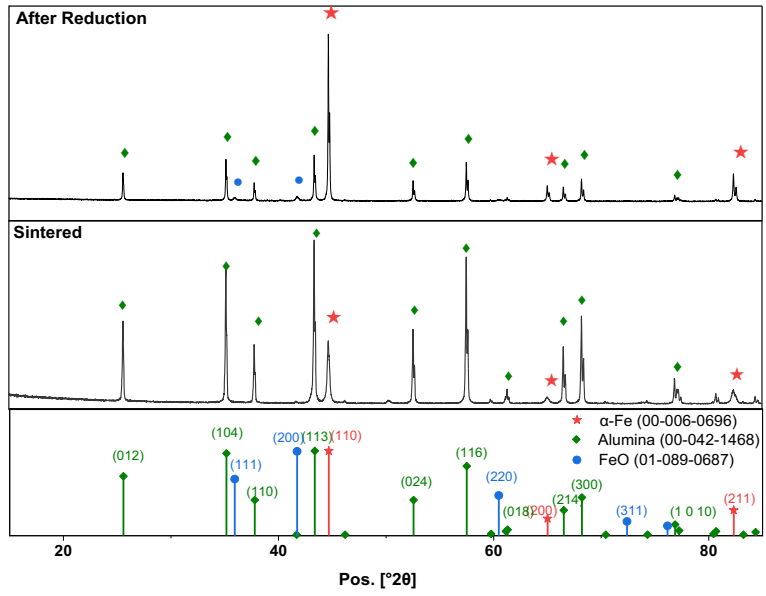

*Figure S5. XRD pattern of sample F-30-F-30 measured on green body, reduced sample (36-hr), and sintered sample in powder form,*